\documentclass[aps,pre,twocolumn,superscriptaddress]{revtex4-1}

\usepackage{amsmath,amssymb,graphicx,textcomp}

\usepackage{times}
\usepackage{color}

\newcommand{\tauf}{\tau_{\text{f}}}
\newcommand{\taup}{\tau_{\text{p}}}

\newcommand{\tauk}{\tau_{\text{k}}}

\newcommand{\beq}{\begin{equation}}
\newcommand{\eeq}{\end{equation}}
\newcommand{\beqa}{\begin{eqnarray}}
\newcommand{\eeqa}{\end{eqnarray}}

\begin{document}
\bibliographystyle{apsrev}

\title{High-resolution detection of Brownian motion for quantitative Optical Tweezers experiments}

\author{Matthias Grimm,$^{1,2}$ Thomas Franosch,$^3$ and Sylvia Jeney$^{*}$}

\address{Laboratory of Physics of Complex Matter, Ecole Polytechnique F\'{e}d\'{e}rale de Lausanne (EPFL), CH-1015 Lausanne, Switzerland\\
$^2$Biozentrum, University of Basel, Klingelbergstrasse 70, Basel 4056, Switzerland\\
$^3$Institut f\"{u}r Theoretische Physik, Friedrich-Alexander-Universit\"{a}t Erlangen-N\"{u}rnberg, Staudtstra{\ss}e 7, 91058 Erlangen, Germany\\
electronic adress: sylvia.jeney@epfl.ch}

\date{\today}
\begin{abstract}
We have developed a new \emph{in situ} method to calibrate optical tweezers experiments and simultaneously measure the size of the trapped particle or the viscosity of the surrounding fluid. 
The positional fluctuations of the trapped particle are recorded with a high-bandwidth photodetector. Next, we compute the mean-square displacement, as well as the velocity autocorrelation 
function of the sphere and compare it to the theory of Brownian motion including hydrodynamic memory effects. A careful measurement and analysis of the time scales characterizing the 
dynamics of the harmonically bound sphere fluctuating in a viscous medium then directly yields all relevant parameters. Finally, we test the method for different optical trap strengths, with different bead
 sizes and in different fluids, and we find excellent agreement with the values provided by the manufacturers. The proposed approach overcomes the most commonly encountered limitations in precision when 
analyzing the power spectrum of position fluctuations in the region around the corner frequency. These low frequencies are usually prone to errors due to drift, limitations in the detection and 
trap linearity as well as short acquisition times resulting in poor statistics. Furthermore, the strategy can be generalized to Brownian motion in more complex environments, provided the adequate 
theories are available.  
\end{abstract}

\maketitle

\section{Introduction}

Single Gaussian beam optical tweezers are typically used to hold and manipulate a single micron-sized sphere in a fluid 
~\cite{Greenleaf2007,Moffitt2008,Veigel2011}.
The sphere fluctuates by thermal excitations within the confines of the harmonic optical trapping potential and is employed as a probe of its local environment 
~\cite{Mason1997,Gisler1998,Furst2005,Wei2010}.
The positional fluctuations are usually recorded by a photodiode, which registers the interference pattern between the trapping laser scattered by the sphere and the unscattered light 
~\cite{Gittes1998,Pralle1999}.
Relying on the linearity of the detector response 
\cite{Jeney2010},
 the position signal measured in Volts is converted into distances by fitting the data to the theory of Brownian motion
~\cite{Svoboda1993,Allersma1998,BergSorensen2004}. 
In its simplest form the Einstein-Ornstein-Uhlenbeck model describes the thermal fluctuations of a sphere in a harmonic potential, 
where the interaction with the surrouding fluid is modelled as an instantaneous Stokes friction. In frequency space, the power spectrum of positional fluctuations is then represented by an overdamped harmonic oscillator with a known viscous drag 
$\gamma=6 \pi \eta R$, where $\eta$ is the fluid's viscosity and $R$ the known bead radius. Fitting the power spectral density (PSD) yields the Volts-to-meter conversion factor, $\beta$, 
and the trap's spring constant $K$ via the corner frequency $f_c$ of the Lorentzian. This straightforward and popular calibration method has been extended in the past years, to allow also for data 
calibration with spheres of unknown sizes or experiencing an unknown viscous drag 
~\cite{Buosciolo2004,Vermeulen2006,Tolic2006,Schaffer2007,Guzman2008,LeGall2011}. 
The common strategy has been
 to add to the intrinsic thermal fluctuations of the bead, an oscillation of known amplitude and frequency by actively moving the fluid or the trap relative to the bead, employing a piezo-stage
 or acousto-optic deflectors respectively.\\
With the improvement of detection bandwidth
~\cite{Peterman2003}, 
also short timescales of optically confined Brownian motion have become accessible 
~\cite{Jeney2008,Huang2011,Franosch2011}.
In principle, three timescales characterize the Brownian motion of a bead held in a harmonic potential. The longest corresponds to the time during which the particle moves back towards the
trap center, $\tau_{\text{K}}=6\pi\eta R/K= 1/2\pi f_c$.\! The Brownian particle displaces the fluid mass $m_{\text{f}}=4\pi \rho_{\text{f}} R^3/3$,
 with $\rho_{\text{f}}$ the fluid density.\! The coupling between the particle and the fluid gives rise to the hydrodynamic memory effect created by a fluid vortex,
which builds up around the particle and diffuses over the distance of one bead radius within the time $\tau_\text{f}=\rho_{\text{f}}R^2/\eta$. 
A third timescale $\tau_\text{p}=M/6\pi\eta R$ is obtained by comparing the stationary Stokes friction with the inertia of the particles.  Here, the effective mass $M=m_{\text{p}}+m_{\text{f}}/2$
results from a calculation including hydrodynamic memory effects.\\
We propose to exploit the nature of hydrodynamic interactions to calibrate high-bandwidth optical tweezers experiments beyond the Einstein-Ornstein-Uhlenbeck theory 
~\cite{BergSorensen2004,Vladimirsky1945,Hinch1975,Clercx1992}. The characteristic timescales of Brownian motion, which are exclusively determined by $R$, $\eta$, $K$, $\rho_{\text{f}}$, and the particle density $\rho_{\text{p}}$, can be directly inferred from the velocity correlation function (VACF) and the mean-square displacement (MSD) of the bead's fluctuations~\cite{Lukic2007}.
The presented method hence does not require a fit to the data rather it relies on reading off characteristic features of the measured curves.\\
In the following, we first give an overview of the experimental requirements. Then, we demonstrate that the MSD and VACF are indeed only dependent on the characteristic timescales of confined Brownian motion. 
Next, we detail the calibration procedure on a representative dataset, and finally test the method for experiments performed with different trap stiffnesses, different bead sizes and different fluids.

\section{Materials and Methods}

The minimal requirements on instrumentation
~\cite{Neuman2004}
 for quantitative optical trapping experiments are: (i) a stable light microscope equipped with a high numerical aperture (NA) 
objective for simultaneous optical visualization and trapping of the probing sphere, (ii) a highly stable trapping laser with an expanded beam for slight overillumination of the objective's back 
aperture, (iii) a photodiode optimally sensitive to the wavelength of the trapping laser for high-resolution and time-resolved position detection
~\cite{Peterman2003}, (iv) high-bandwidth signal 
amplification, and (v) a data acquisition board that allows for a sampling rate of at least 1 MHz.\\
\noindent It is also convenient to precisely control by, for example a three-dimensional (3D) piezo stage, the distance between the sphere under study and boundaries in the sample chamber
~\cite{Franosch2009,Jeney2010}.\\
\noindent More specifically, our instrument includes a diode-pumped, Gaussian beam, ultra-low noise Nd:YAG laser with a wavelength of $\lambda$=1064 nm, and a maximal light power of 500 mW in
 continuous wave mode (IRCL-500-1064-S, CrystaLaser, USA). Best trapping efficiency is achieved by expanding the effective laser beam diameter 10 times with a telecentric lens system 
(Sill Optics, Germany).
The IR-beam is reflected by a dichroic mirror (AHF Analysentechnik AG, Germany) into the high numerical aperture (NA=1.2) of a 60x water-immersion objective (UPLapo/IR, Olympus, Japan), which
 focuses the laser down to its diffraction limit into the object plane of the microscope and creates the optical trap. The resulting beam waist, $w_0$, is approximately 118 nm in water with a 
refractive index n=1.32 at $\tiny \lambda$=1064 nm
~\cite{Nieminen2007}. The choice of a water-immersion objective lens offers a longer working distance of up to 280 \textmu m compared to oil-immersion
 lenses, and guarantees a stable uniform trap through the entire sample chamber. The sample is mounted onto an xyz-piezo scanning table (P-561, Physikalische Instrumente, Germany) for 
manipulation and positioning. The piezo-stage with controller (E-710 Digital PZT Controller, Physikalische Instrumente, Germany) has a travel range of 100 \textmu m in 3D with a precision 
of $\sim$1 nm.
The laser light focused by the objective lens is collected with a second objective lens (63X, Achroplan, NA = 0.9, water-immersion, Zeiss, Germany), 
and focussed onto an InGaAs quadrant photodiode with an active area of 2 mm in diameter (QPD, G6849, Hamamatsu Photonics, Japan). The QPD is fixed to a x-y translation stage (OWIS, Germany) 
for manual centering of the detector relative to the IR-beam.\\
The QPD signals are fed into a pre-amplifier (\"{O}ffner MSR-Technik, Germany), which provides two differential signals between the segments and one signal that is proportional to the total 
light intensity. This allows the detection of the particle's position in 3D. Pre-amplification of the QPD signals at 20 V/mA with 0.67 A/W photosensitivity leads to a voltage of 13.4 V/mW. Subsequently, 
differential amplifiers (\"{O}ffner MSR-Technik, Germany) adjust the pre-amplifier signals for optimal digitalization by the data acquisition board with a dynamic range of 12 bits (NI-6110, 
National Instruments, USA). In accordance with the total laser intensity impinging on the QPD, amplification of the QPD signal is chosen to span the maximal dynamic range of the acquisition card.
 The amplifier, with a maximal gain of 500, has a cut-off frequency around 1 MHz. \\
The sample chamber consists of a custom-made flow cell. A coverslip (thickness $\sim 130$ \textmu m) is glued to a standard microscope slide by two pieces of double-sided tape arranged to form a $\sim 5$ mm wide
 and $\sim 80$ \textmu m thick channel with a volume of $\sim 30$ \textmu l. After loading with a dilute suspension of microspheres, the flow-cell is mounted upside down on the 3D piezo-stage of our custom-made inverted microscope.\\
To achieve the high temporal resolution required for the present calibration method, it is crucial to choose the detector material to be optimally sensitive to the wavelength of the
 trapping laser 
~\cite{Peterman2003}
, which in our case is InGaAs. Furthermore, a smaller active area results in a lower dark current and a higher signal-to-noise ratio at short times.
 For 3D position detection, we chose a QPD diameter of 2 mm onto which we focussed the trapping laser spot. An optimally focussed laser spot yielded a temporal resolution close to 
1 \textmu s. If 3D position detection is not crucial, even higher temporal resolution down to 10 ns in a single spatial direction can be achieved using balanced amplified photodetectors with two 
well-matched photodiodes with a diameter of 0.3 mm each, and an ultra-low noise, high-speed transimpedance amplifier ~\cite{Huang2011}.\\

\section{Brownian Motion}
The individual trajectory of a particle undergoing Brownian motion is in general unpredictable. Nevertheless, correlation functions such as the velocity autocorrelation function $\langle v(t)v(0)\rangle$  and the 
mean-square displacement $\langle\Delta x^2(t)\rangle=\langle [x(t)-x(0)]^2\rangle$ are theoretically
accessible and describe the statistical properties of the dynamics. Both functions are connected via $\langle v(t)v(0)\rangle=(1/2)\text{d}^2 \langle\Delta x^2(t)\rangle/\text{d} t^2 $.
 The displacement of the particle relative to the trap center as a function of time, $x(t)$, obeys the Langevin equation
\begin{equation}
\label{Langevin}
  m_{\text{p}}\ddot x(t)=F_{\text{th}}(t)+F_{\text{fr}}(t)-Kx(t)\,,
\end{equation}
where $m_{\text{p}}=4\pi\rho_{\text{p}}R^3/3$ corresponds to the particle's mass, $F_{\text{th}}$ denotes the thermal random noise and $F_{\text{fr}}$ the friction forces acting on the particle. The friction force, known as Basset force, consists of Stokes' law and 
additional terms accounting for
the fluid vortex developing around the particle
\begin{eqnarray}
 F_{\text{fr}}(t)=&-&6\pi\eta R\dot x(t)-\frac{2}{3}\pi R^3\rho_{\text{f}}\ddot x(t)\nonumber\\
&-&6R^2\sqrt{\pi\rho_{\text{f}}\eta} \int_{-\infty}^t\frac{\ddot x(t')}{\sqrt{t-t'}}\,\text{d}t'\,,
\end{eqnarray}
and is valid for spherical particles with no-slip boundary conditions on their surface \cite{Landau1959}.
Here, the last contribution depends on the entire history of the particle's motion and therefore it is referred to as ``hydrodynamic memory''. \\
The thermal force describes the random collisions of the surrounding solvent molecules
with the particle. While its mean vanishes by symmetry arguments, its autocorrelation is given by
\begin{equation}
 \langle F_{\text{th}}(t) F_\text{th}(t')\rangle =  k_B T \gamma(|t-t'|)\,,
\end{equation}
where $\gamma(t)$ is a kernel describing the retarded friction forces.
The explicit form of $\gamma(t)$  requires the use of distributions, but its Laplace transform (convention $\hat x (s)=\int_{0}^{\infty} x(t)\exp(-st)\text{d}t$) takes the following form 
\cite{Franosch2011,Landau1959,Kubo1991}
\begin{equation}
 \hat{\gamma}(s) = 6\pi \eta R [ 1 + \sqrt{s \tau_\text{f} } ]  + s m_\text{f}/2\,. 
\end{equation}
Here, the constant describes the delta-correlated white noise usually assumed for Brownian motion. The last term can be intrepreted as added mass contibuting to the effective mass $M$, whereas the non-analytic square-root accounts for the hydrodynamic memory.\\
From the Langevin equation one derives the VACF in the Laplace domain, normalized to its initial value $\langle v(0) v(0)\rangle=k_B T/M$, as 
\begin{equation}
\label{VAFs}
 \frac{\widehat{\langle v(s) v(0)\rangle}}{\langle v(0) v(0)\rangle}=\frac{s}{P(s)}\,,
\end{equation}
with the polynomial $P(s)=s^2+s(1+\sqrt{s\tauf})/\taup+1/(\taup\tauk)$. 
The inverse Laplace transform is achieved observing that the Laplace transform of $\exp(z^2 t/\tauf)\text{erfc}(z\sqrt{t/\tauf})$ is $1/(s+z\sqrt{s/\tauf})$. 
The VACF can be evaluated by partial fraction decomposition of Eq.~(\ref{VAFs})
\begin{equation}
 \frac{\widehat{\langle v(s) v(0)\rangle}}{\langle v(0) v(0)\rangle}=\sum_{j=1}^4 \frac{A_j}{s+z_j\sqrt{s/\tauf}}\,,
\end{equation}
where $A_j=\lim_{\sqrt{s}\to - z_j / \sqrt{\tau_f } } s (s+ z_j \sqrt{s/\tau_f} )/P(s)=\prod_{k\neq j}[z_j/(z_j-z_k)]$ and the $z_j$ are the four roots of the polynomial $P(s)=\prod_{j=1}^4 (  \sqrt{s} + z_j /\sqrt{\tauf}) $. Finally, the normalized VACF is given by
\begin{equation}\label{eq:VAF}
\frac{{\langle v(t) v(0)\rangle}}{\langle v(0) v(0)\rangle}=\sum_{j=1}^4 A_j \exp(z_j^2 t/\tauf)\text{erfc}(z_j\sqrt{t/\tauf})\,. 
\end{equation}
Since in the Laplace domain, $\widehat{\langle v(s)v(0)\rangle} = s^2 \langle\widehat{\Delta x^2(s)}\rangle/2$, a similar partial fraction decomposition may be employed to calculate the MSD.
After normalization to its plateau value, $\langle\Delta x^2(t\to\infty)\rangle=2k_B T/K$, the MSD reads
\begin{eqnarray}\label{eq:MSD}
  \frac{\langle\Delta x^2(t)\rangle}{\langle\Delta x^2(t\to\infty)\rangle}&=&1\nonumber\\+\frac{\tauf^2}{\taup\tauk}\sum_{j=1}^4 (&A_j&/z_j^4) \exp(z_j^2 t/\tauf)\text{erfc}(z_j\sqrt{t/\tauf})\,. 
\end{eqnarray}
Note that Eqs.~(\ref{eq:VAF}) and (\ref{eq:MSD}) only depend on the ratio $t/\tauf$. The trap strength, and the mass of the particle are encoded in the dimensionless roots of the polynomial
$P(s)$ and in the amplitudes $A_j$.\\
The above description is valid in the case of a spherical particle performing Brownian motion in a simple fluid in infinite space~\cite{Clercx1992}. Additional time scales arise in more complex systems.
For example, when approaching a surface, the time scale $\tau_{\textsf{w}}=\rho_{\textsf{f}}h^2/\eta$, with $h$ the distance to the surface, describes the influence of the wall on the Brownian Motion. 
\cite{Jeney2008,Franosch2009}. Furthermore, a viscoelastic fluid has a timescale that marks the transition from its purely viscous to elastic regime~\cite{Grimm2011}. However, in such cases, no analytical expressions for the MSD and VACF are available and their calculation has to rely on numerical
Fourier transforms. In the following section, we show how Brownian motion in an optical trap may be used to calibrate the optical tweezers only in the case where analytic expressions
exist.

\section{Procedure}
Interestingly, as highlighted by Eqs. (\ref{eq:VAF}) and (\ref{eq:MSD}), the shapes of the normalized MSD and VACF are only determined by the ratios $\tau_\text{p}/\tau_\text{f} = (2\rho_\text{p}/9 \rho_\text{f}+1/9)$
and $\tau_\text{K}/\tau_\text{f}=6 \pi \eta^2/\rho_{\text{f}}KR$. Since the temperature and densities of the fluid and the bead are usually known or can be easily determined by other means, the shape of the curves are sufficient to determine the ratio $\tau_\text{K}/\tau_\text{f}$ and the timescale $\tauf$.
Therefore, from reading out $\tau_\text{p}, \tau_\text{f}, \tau_\text{K}$, the scale factor $\beta$ can be inferred. Eventually, the trap stiffness $K$ and the bead radius $R$ or the viscosity $\eta$ follow from the definition of $\tau_\text{f}$ and $\tau_\text{K}$.\\
Fig. 1 displays the uncalibrated $\langle\Delta x^2(t)\rangle^{\textsf{V}}$ and $\langle v(t)v(0)\rangle^{\textsf{V}}$ calculated directly from the position signal of an optically trapped sphere in water measured in Volts during 20 s at a sampling frequency $f_S=1$ MHz.
Assuming a linear detector response 
\cite{Jeney2010}
, with $\beta^{-1}$ the detector sensitivity in $\text{V}/\text{m}$, the actual position of the particle is given by
\begin{equation}
x_j(t)= \beta x^{\textsf{V}}_j(t)\,,\quad j=1...N\,,
\end{equation}
where $x^{\textsf{V}}$ is the voltage measured by the detector for N acquired data points (here $N=2\cdot10^7$). The sampling frequency determines the time interval between two data points 
$\delta t=1/f_S=1\,\text{\textmu s}$. 
The mean-square displacement is calculated from the raw data as:
\begin{equation}
 \langle\Delta x^2(t)\rangle^{\textsf{V}}=\frac{1}{N}\sum_{j=1}^N \left[x^{\textsf{V}}(t+t_j)-x^{\textsf{V}}(t_j)\right]^2\,,
\end{equation}
and the VACF is determined from the increments of the measured voltages through
\begin{eqnarray}
 \langle v(t)v(0)\rangle^{\textsf{V}}=\frac{1}{N\delta t^2}\sum_{j=1}^N \left[x^{\textsf{V}}(t+t_j+\delta t)-x^{\textsf{V}}(t+t_j)\right]\nonumber\\
\times\left[x^{\textsf{V}}(t_j+\delta t)-x^{\textsf{V}}(t_j)\right].
\end{eqnarray}
For calibration, two values can be directly read off the raw MSD data (Fig. 1(a)): first, the long-time plateau $\langle\Delta x^2(\infty)\rangle^{\textsf{V}}$, which allows the normalization of the data (inset), 
and second, the time $t_{1/2}$ at which $\langle\Delta x^2(t)\rangle^{\textsf{V}}$ reaches half of its maximum. The MSD is essentially the Fourier transform of the PSD, most commonly used for the calibration of optical
 traps in the frequency domain 
~\cite{Svoboda1993,Allersma1998,BergSorensen2004}. Similarly, the VACF graph displays two readily accessible characteristic features (Fig. 1(b)): the time $t_0$ at
 which it crosses zero and becomes negative, and the value of its minimum $\langle v(t_{min})v(0)\rangle^{\textsf{V}}$. The anticorrelations in the VACF originate from the harmonic restoring force of the optical trap and are discussed later in Fig. 3(b). The earlier they occur, the stronger the trap.
For weaker traps, a third interesting feature appears in the positive short-time values of the VACF, which decays with a $t^{-3/2}$ power-law (Fig. 1(b), inset). The amplitude of this well-known long-time tail (LTT) is given by $B=k_B T\sqrt{\rho_{\text{f}}}/12(\pi\eta)^{3/2}$ and depends only on the fluid properties \cite{Jeney2008,Hinch1975}.\\

\begin{figure*}\centering
\includegraphics[scale=0.08]{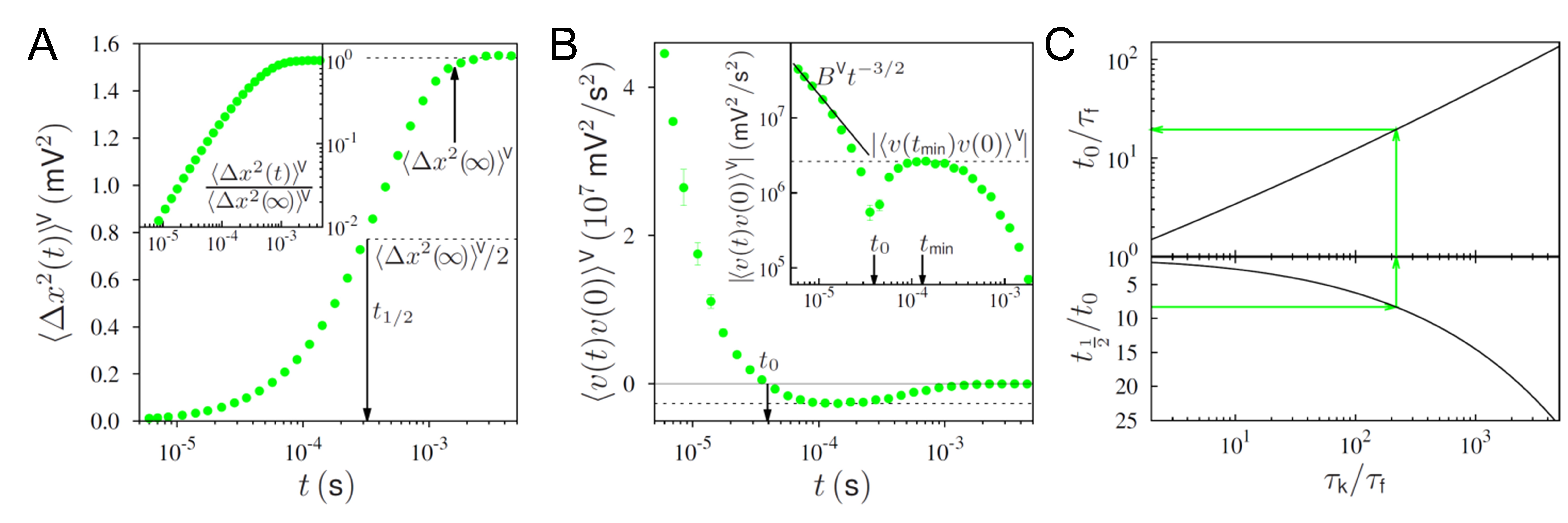}%0.06
\caption{(Color online) (a) Lin-log representation of the uncalibrated MSD of an optically trapped melamine resin microsphere in water
($R=1.4$\,\textmu m, $\rho_{\text{p}}=1510$ g/l, $\rho_{\text{f}}=1000$ g/l, $\tau_\text{p}/\tau_\text{f}=0.45$) yielding  $\langle\Delta x^2(\infty)\rangle^{\textsf{V}}=1.54\, mV^2$ and t$_{1/2}=327$\,\textmu s. 
 Inset: Log-log representation of the same MSD normalized by its plateau value. (b) Lin-log representation of the corresponding uncalibrated VACF, with $t_0=39.3$ \textmu s and 
$\langle v(t_{min})v(0)\rangle^{\textsf{V}}=-2.64\cdot 10^6\,\text{mV}^2/\text{s}^2$.
Inset: Log-log representation of $|\langle v(t)v(0)\rangle^{\textsf{V}}|$. The amplitude of the $t^{-3/2}$ long-time tail reads $B^{\textsf{V}} =0.65 \text{mV}^2/\text{s}^{1/2}$. Data were blocked in 10 bins per decade. Error bars give the standard error on the mean from blocking. 
(c) Top: Theoretical ratios t$_{1/2}/\tau_\text{f}$  and t$_{0}/\tau_\text{f}$ as a function of $\tauf/\tauk$ for a resin sphere ($\rho_{\text{p}}=1510$ g/l) in water ($\rho_{\text{f}}=1000$ g/l). The trap relaxation time increases from left to right, while the trap stiffness decreases. Bottom: Corresponding theoretical values of t$_{1/2}/t_0$.
 Each value necessary to calibrate the data shown as an example is highlighted by an arrow.
}\label{fig1}
\end{figure*}

\noindent The ratio $t_{1/2}/t_0$ determined from the experimental data uniquely sets $\tau_\text{K}/\tau_\text{f}$. However, at this step, a comparison between the theory and the data in Fig. 1 requires fitting with at least 3 variables, $\beta$, $K$ and $R$ or $\eta$.
To reduce the number of fit parameters, we compute the theoretical functions given by Eqs. (\ref{eq:VAF}) and (\ref{eq:MSD}) for a wide range of ratios of $\tau_\text{K}/\tau_\text{f}$, but fixed $\tau_\text{p}/\tau_\text{f}$, hence a given particle and fluid density.\\
\noindent
The theoretically possible values for $t_{1/2}$ as a function of $\tauk/\tauf$ are then determined from $\langle\Delta x^2(t_{1/2})\rangle/\langle\Delta x^2(t\to\infty)\rangle=1/2$,
and $t_{0}$ is given by $\langle v(t_0)v(0)\rangle/\langle v(0)v(0)\rangle=0$. Consequently, the value of $\tauk/\tauf$ corresponding to an experimentally measured value of $t_{1/2}/t_0$ can be 
directly read off from the bottom graph in Fig. 1(c). Once the ratio $\tauk/\tauf$ is known, the values for $t_{1/2}/\tau_\text{f}$ and $t_{0}/\tau_\text{f}$ are taken from the respective graphs on top in Fig. 1(c), allowing the inference of $\tauf$, and subsequently $\tauk$, $K$ and either $R$ or $\eta$.\\

\noindent
Finally the conversion factor $\beta$ follows by either: (i) overlaying the theoretical and experimental plateau values of the MSD: $\beta_{\textsf{MSD}}^2=2k_B T/(K\langle\Delta x^2(\infty)\rangle^{\textsf{V}})$,
(ii) matching the minimum value $\langle v(t_{min})v(0)\rangle$ of the experimental data to the minimum of the normalized theoretical curve: $\beta_{\textsf{VACF}}^2=\langle v(t_{min})v(0)\rangle/\langle v(t_{min})v(0)\rangle^{\textsf{V}}$, or (iii) matching the amplitude of the LTT measured in V$/s^{1/2}$ to its theoretical value: $\beta_{\textsf{LTT}}^2=B/B^{\textsf{V}}$.\\
\noindent
Applying the described calibration procedure to the exemple dataset presented in Fig. 1 results in the following steps:
$t_{1/2}/t_0=8.1 \rightarrow \tau_\text{K}/\tau_\text{f}=219\quad\text{(Fig. 1(c), bottom)} \rightarrow t_{0}/\tauf=19.7\quad\text{and}\quad t_{1/2}/\tauf=162\quad\text{(Fig. 1(c), top)} 
\rightarrow \tauf=2\,\text{\textmu s} \rightarrow \tauk=440\,\text{\textmu s}$ $\rightarrow$ for water with $\eta=0.98\,\text{cP}$ at $T=21^\circ\,\text{C}$, we obtain 
$R=\sqrt{\eta\tauf/\rho_{\text{f}}}=1.4\, \text{\textmu m}$ and $K=6\pi\rho_{\text{f}} R^3/(\tauf\tauk)=59\, \text{\textmu N/m}=0.059\, \text{pN/nm}$, yielding three values for the calibration factor
 $\beta$: $\beta_{\textsf{MSD}}=9.4$\,nm/V, $\beta_{\textsf{VACF}}=8.8$\,nm/V, and $\beta_{\textsf{LLT}}=8.8$\,nm/V. A fit to the PSD as discussed in ref. \cite{BergSorensen2004}
results in $\beta_{\textsf{PSD}}=9.8$\,nm/V (data not shown), in good agreement with the three values computed by the method presented here. However, a reliable fit of the PSD can only be achieved with two unknown parameters, typically K and beta. Reading off characteristic features from the data rather than fitting to the data allows to determine a third parameter.
\noindent
Possible discrepancies between the three, differently obtained $\beta$-values and $\beta_{\textsf{PSD}}$ point to the limits in the reading accuracy of each characteristic feature,
 as well as instrumental limitations such as mechanical drift and electronic noise. They are discussed in the next section on measurements with different trap stiffnesses, radii and viscosities.

\section{Tests}
\noindent
\begin{figure*}\centering
\includegraphics[scale=0.08]{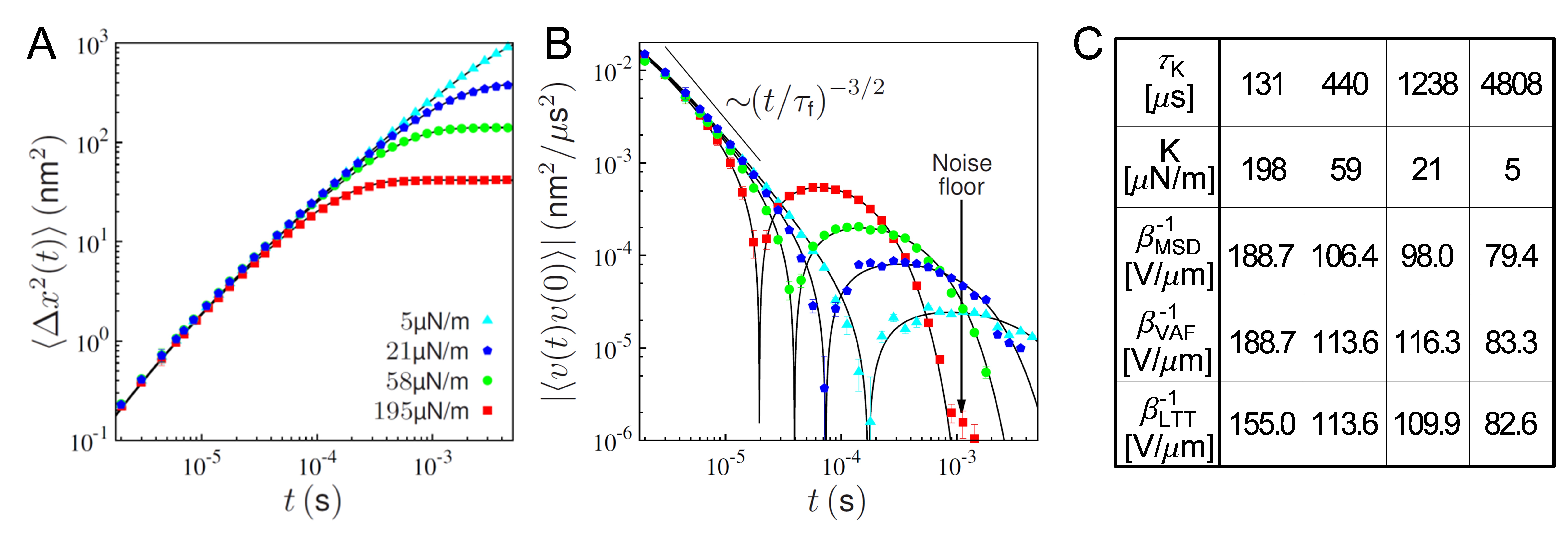}
\caption{(Color online) Calibrated MSD (a) and VACF (b) of single resin sphere ($\rho_{\text{p}}=1510\,$g/l, $R=1.4\, \text{\textmu m}$) in water ($\rho_{\text{f}}=1000\,$g/l, $\eta=0.98\,\text{cP}$, $\text{T}=21^{\circ}\,\text{C}$, $\taup/\tauf=0.45$, 
$\tauf=2\,\text{\textmu s}$) for increasing $K$. Data were blocked in 10 bins per decade. Error bars give the standard error on the mean from blocking. (c) Overview of the parameters obtained from calibrating the data.}
\label{fig3}
\end{figure*}
\textbf{Changing the trap stiffness:} Fig. 2 displays the calibrated MSDs and VACFs of the same resin sphere immersed in water, for various $K$. 
The trap stiffness was increased by increasing the incoming trapping laser power. The black lines give the corresponding theoretical predictions.
As expected, all curves display the same long-time tail, and the plateau of the MSD as well as the zero-crossing in the VACF appear the earlier given a stronger trap and hence shorter $\tauk$.
 Also the minimum of the VACF is deeper, the higher $K$. In contrast, the LTT, which is influenced by the fluid's density and viscosity, disappears from our measurement window, 
as motion becomes dominated by the trap. As a consequence, depending on the experimental conditions, the features of the MSD and VACF may be more or less pronounced. 
In general, the plateau in the MSD and $t_{1/2}$ can be best determined the earlier they are reached. In weaker traps, $\langle\Delta x^2(\infty)\rangle^{\textsf{V}}$ occurs later and may be noisier due to mechanical drift.
 For optimal readability, the zero-crossing in the VACF has to occur at $t_0 \geq 2\,\text{\textmu s}$, above the resolution limit of our detector. 
Furthermore, $\langle v(t_{min})v(0)\rangle^{\textsf{V}}$ has to be pronounced enough to be distinguishable from the electronic noise floor, which here is around $10^6\,\text{nm}^2/\text{\textmu s}^2$. 
Such conditions are fulfilled for $1\, \text{\textmu N/m}\lesssim K \lesssim 500\, \text{\textmu N/m}$, the typical stiffness range of optical traps. When $\tauk \gtrsim 500\,\text{\textmu s}$ 
a clearly visible long-time tail exists and $B^{\textsf{V}}$ can in turn be evaluated accurately. The parameters $K$ and the detector sensitivity $\beta^{-1}$ obtained by the present method are listed in Fig. 2(c). The highest sensitivity and the best match between $\beta^{-1}_{\textsf{MSD}}$ and $\beta^{-1}_{\textsf{VACF}}$ is obtained for strong traps, when all respective features are easily distinguishable.\\
\noindent
\begin{figure*}
\centering
\includegraphics[scale=0.08]{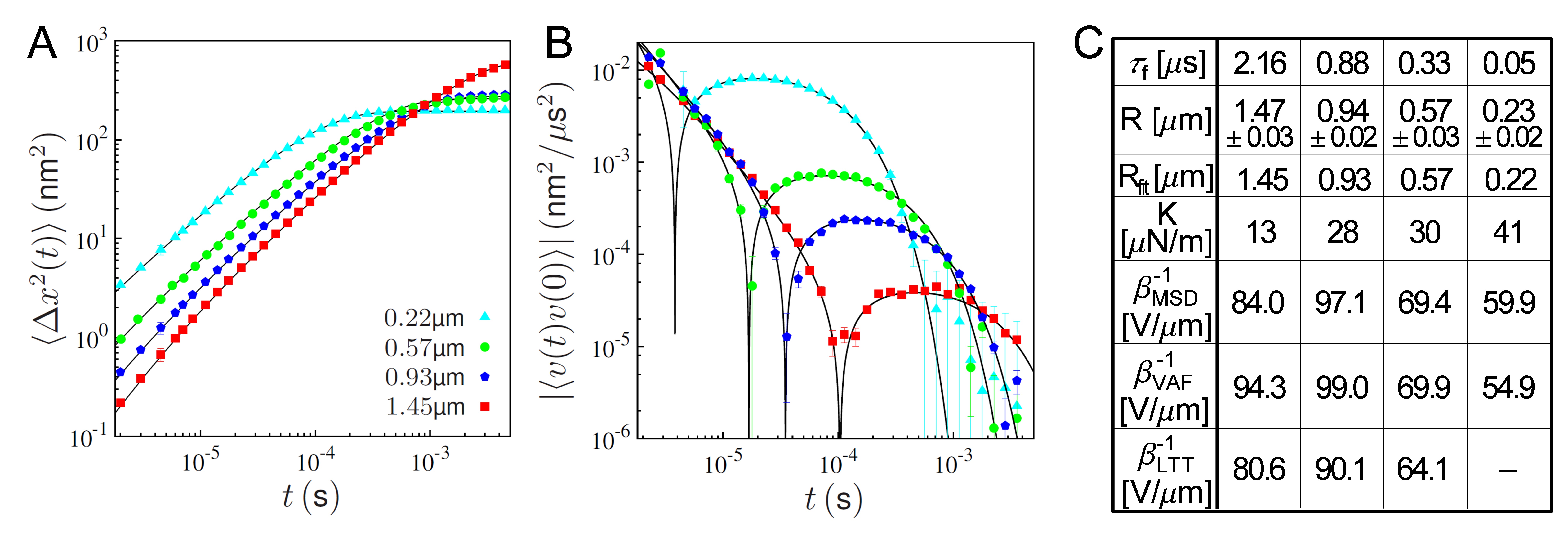}
\caption{(Color online) Calibrated MSD (a) and VACF (b) of single resin particles ($\rho_{\text{p}}=1510\text{g/l}$) with different sizes in water 
($\rho_{\text{f}}=1000\text{g/l}$, $\eta=0.98\,\text{cP}$, $\taup/\tauf=0.45$). Data were blocked in 10 bins per decade. Error bars give the standard error on the mean from blocking. 
(c) Overview of the parameters obtained from calibrating the data.}
\label{fig5}
\end{figure*}
\textbf{Changing the particle size:} Next, we decreased $R$ and adapted the laser power in order to have approximately similar trap stiffnesses for each bead (Fig. 3). 
Changing the bead size influences the measured signal in an non-linear way and at several levels. First, and most straightforwardly, hydrodynamic backflow is enhanced with $R^2$ via $\tauf$,
 while confinement occurs earlier for bigger particles through $\tauk$. Second, the ratio between bead size and laser wavelength, very strongly influences the optical forces acting on the particle.
 Empirically and according to theory, forces are strongest for $2R \simeq\lambda/n\simeq 800$\,nm 
~\cite{Rohrbach2005}
. Third, the ratio $R/w_0$ determines the total light intensity impinging on
 the detector, as well as the interference pattern between scattered and non-scattered light. These parameters influence the detector sensitivity $\beta^{-1}$, as well as the noise floor of the 
signal, which appears to be higher for smaller beads (Fig. 3). Apart from determining $K$ and $\beta^{-1}$, we also inferred $R$ directly from the data. All obtained parameters are given in the
 table of Fig. 3, showing that the detector sensitivity is reduced for $2R\lesssim\lambda$. The obtained $R_{\text{fit}}$-values for an individual particle are in excellent agreement with 
the bead sizes, $R$ provided by the manufacturer. However, for bead radii yielding $\tau_\text{f}=\rho_{\text{f}}R^2/\eta <1\,\text{\textmu s}$, hydrodynamic effects become less prominent
in our measurement window, which impairs the reading off accuracy of the short-time features, $B^{\textsf{V}}$ and $t_{0}$. A simultaneous quantification of $R$ and $K$ then becomes a challenge. Here, a detector with an improved temporal resolution as used in 
\cite{Huang2011} 
would allow to measure even smaller particle sizes.\\
\noindent
\begin{figure*}
\centering
\includegraphics[scale=0.08]{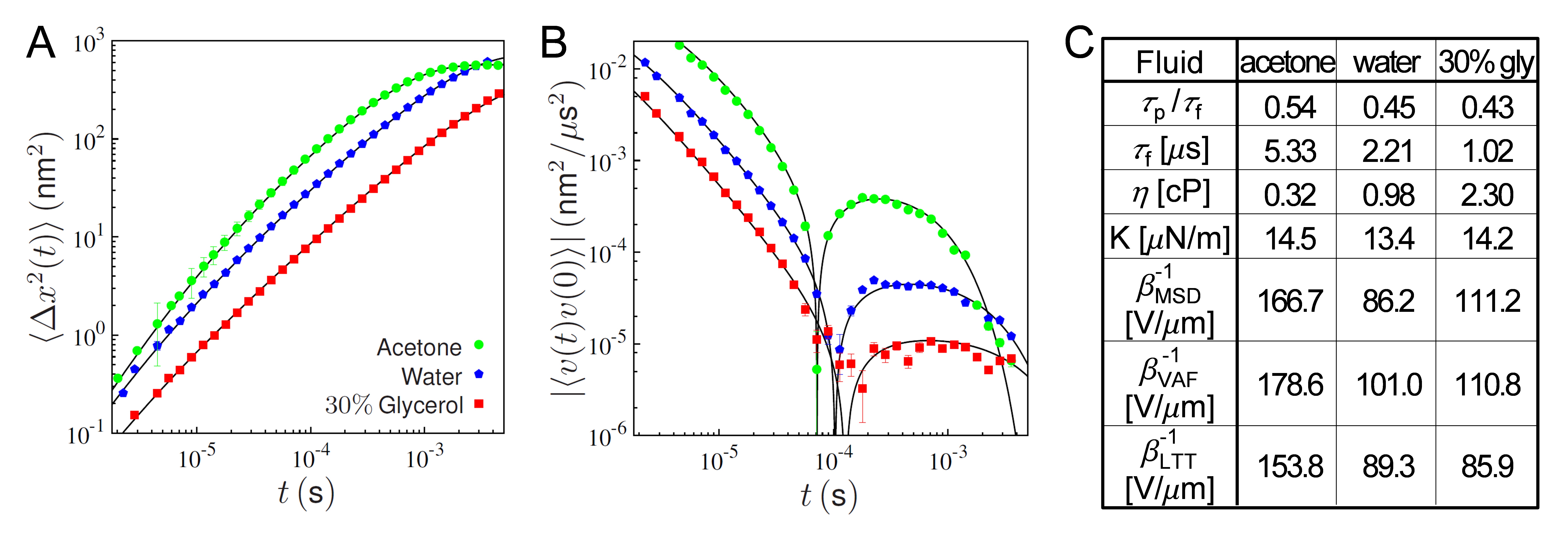}
\caption{(Color online) Calibrated MSD (a) and VACF (b) of resin particles in acetone ($\rho_{\text{f}}=790\,\text{g/l}$),
 water ($\rho_{\text{f}}=1000\,\text{g/l}$)
and glycerol ($\rho_{\text{f}}=1066\,\text{g/l}$). Data were blocked in 10 bins per decade. Error bars give the standard error on the mean from blocking. (c) Overview of the parameters obtained from calibrating the data.}
\label{fig4}
\end{figure*}
\textbf{Changing the fluid viscosity:} Finally, we tested our method to calibrate the optical trap and simultaneously measure the viscosity of a fluid, when the bead size is known and the
 laser power is held constant. Different fluids also 
alter the measured signal in a non-linear way. Firstly, the refractive index and the absorption coefficient of the medium influence the trapping forces and the light pattern reaching the detector. 
Secondly, a higher viscosity will yield weaker trap stiffnesses as quantified by $\tauk$. Finally, the shape of the VACF (Eq. (\ref{eq:VAF})) and the MSD (Eq.(\ref{eq:MSD})) are most dependant on $\eta^2$ through 
 $\tauk/\tauf$. In contrast, a shift of $\taup/\tauf$ corresponding to a change in the fluid density has a negligible effect on the MSD and VACF. Fig. 4 displays the calibrated MSD (a) and VACF 
(b) of a resin sphere ($R=1.47\,\text{\textmu m}$) immersed either (i) in the low viscosity solvent acetone, (ii) 
in water, or (iii) in $30\%$ glycerol, which is 2 to 3 times more viscous than water. The trap stiffness $K$, the viscosity $\eta$ and the conversion factor $\beta^{-1}$ were determined from 
the presented method and are listed in Fig. 4(c).  
The obtained fluid viscosities are in good
agreement with viscosimetry measurements, which valued $\eta=0.32\,\text{cP}$\, for aceton, $\eta=0.98\,\text{cP}$ for water, and $\eta=2.3\,\text{cP}$ for 30\% glycerol in water at ambient temperature. 

\section{Discussion and Conclusion}
In summary, we have established a procedure to calibrate optical trapping experiments with a micron-sized particle immersed in a viscous fluid. The method allows the determination of the position detector sensitivity, the trap stiffness and simultaneously the particle size or the fluid viscosity with high precision. When neither $\eta$ nor $R$ are known, we suggest performing two consecutive measurements at two different trap strengths. For a weak trap, $\eta$ can be extracted from the short-time hydrodynamic features given by the amplitude $B$ of the $t^{-3/2}$ power-law. 
Upon knowing the fluid properties, the procedure can be pursued as described.
Furthermore, the presented method can be applied equivalently in both directions, x and y, perpendicular to the optical axis, yielding $K_x$, $K_y$, $\beta_x$ and $\beta_y$. For the dimension,
 z along the optical axis, the signal-to-noise ratio, and hence bandwidth of the position detector is usually not sufficient to measure a smooth VACF exhibiting a clearly detectable zero-crossing 
(see supplemenary information in 
\cite{Franosch2011}). Nevertheless, after having determined the bead radius and/or fluid viscosity through analysis of the x or y dimension, $\beta_z$ and $K_z$ 
can be straightforwardly inferred from the plateau value of the MSD and the corner frequency $f_c=1/2\pi \tauk$ of the PSD 
\cite{BergSorensen2004}, which is much less prone to 
high-frequency noise than the VACF. Measuring in 3D should allow the assesment of anisotropies in the bead sizes as well as in the surrounding medium. 
The proposed approach should be applicable in 
microrheology for the study of viscoelastic fluids, in which at least one additional timescale marks the transition between purely viscous and elastic behavior \cite{Grimm2011}. Also, when the
Brownian particle approaches a surface, a new characteristic time can be extracted from the data, indicating the distance between the particle and the surface \cite{Jeney2008}.
\\
For example, when the Brownian particle approaches a surface, the time needed by the hydrodynamic vortex to diffuse from the particle to the surface can be read from the data. Measuring this 
time yields the distance between the particle and the surface 
 between a purely viscous and an 
elastic behavior of the fluid 
\cite{Grimm2011}, and characterizes Brownian motion.

\section{ACKNOWLEDGEMENTS}
M.G. is supported by NCCR Nanoscale Science. S.J. acknowledges the Swiss National Science Foundation (SNF grant nos 200021-113529 and 206021-121396). We thank Flavio M. Mor for discussions.

\bibliographystyle{apsrev}

\end{document}